\newcommand{\diracslash}[1]{#1\llap{/\kern2pt}}

\newcommand{\be}{\begin{equation}}
\newcommand{\ee}{\end{equation}}
\newcommand{\bea}{\begin{eqnarray}}
\newcommand{\eea}{\end{eqnarray}}
\newcommand{\ba}[1]{\begin{array}{#1}}
\newcommand{\ea}{\end{array}}

\newcommand{\bt}{\begin{tabular}}
\newcommand{\et}{\end{tabular}}

\newcommand{\beas}{\begin{eqnarray*}}
\newcommand{\eeas}{\end{eqnarray*}}

\documentclass[preprint,prd,aps,amssymb,amsmath
,floats,nofootinbib,floatfix]{revtex4}

\DeclareSymbolFont{rsfs}{U}{rsfs}{m}{n}
\DeclareSymbolFontAlphabet{\mathrsfs}{rsfs}

\usepackage{graphicx}
\usepackage{multirow}
\usepackage{graphicx,epstopdf}

\usepackage{graphicx}
\usepackage{float}

\usepackage[utf8]{inputenc}
\usepackage[english]{babel}
\usepackage{hyperref}
\hypersetup{
    colorlinks=true,
    linkcolor=blue,
    filecolor=magenta,     
    citecolor=blue, 
    urlcolor=blue,
}
 
\usepackage{cleveref}

\begin{document}

\title{$\phi$ meson mass and decay width in strange hadronic matter} 

\author{Rajesh Kumar}
\email{rajesh.sism@gmail.com}

\author{Arvind Kumar}
\email{iitd.arvind@gmail.com, kumara@nitj.ac.in}
\affiliation{Department of Physics, Dr. B R Ambedkar National Institute of Technology Jalandhar, 
 Jalandhar -- 144011,Punjab, India}
%

\def\be{\begin{equation}}
\def\ee{\end{equation}}
\def\bearr{\begin{eqnarray}}
\def\eearr{\end{eqnarray}}
\def\zbf#1{{\bf {#1}}}
\def\bfm#1{\mbox{\boldmath $#1$}}
\def\hf{\frac{1}{2}}
\def\kp{\zbf k+\frac{\zbf q}{2}}
\def\km{-\zbf k+\frac{\zbf q}{2}}
\def\hwo{\hat\omega_1}
\def\hwt{\hat\omega_2}

\begin{abstract}
The in-medium mass and decay width of the $\phi$ meson in the hot  asymmetric strange  hadronic matter are calculated using an effective Lagrangian approach for $\phi$ $K\bar K$ interaction. The in-medium contributions of $K \bar K$ loop to the $\phi$ meson self-energy are computed by the medium modified kaon and antikaon mass which is evaluated using the chiral SU(3) model. To deal with the ultraviolet divergence, we regularize  the loop integral of self-energy with a dipole form factor, and  present the results for  cut-off mass, $\Lambda_c$=3 GeV. In chiral model calculations, for strange matter,  we find that the mass of kaons and antikaons decreases with the increase in  baryonic density whereas the finite temperature causes an increase in the mass. We observed that the in-medium $\phi$ meson mass decreases slowly with baryonic density whereas the decay width increases rapidly. The results in the present investigation support result in the literature which indicate a small downward shift in mass and a large broadening in the decay width. In the asymmetric strange hadronic matter, the study of the $\phi$  mesons  can be  relevant
for the compressed strange baryonic matter and experimental observables such as dilepton spectra which can result from the 
experiments in the future FAIR facility at GSI.

\end{abstract}

\maketitle

\maketitle

\section{Introduction}
\label{intro}
The study of the hadron properties  in the strange asymmetric matter at finite temperature is of considerable interest to understand the QGP phase diagram in the strong interaction physics \cite{Tolos2020,Holzenkamp1989,Petschauer2016,Haidenbauer2019,Kumar2015,Chhabra2017,Chhabra2018,Mishra2009,
Kumar2011}. Experimentally, the heavy ion-collisions (HICs) play an important role to study the hot and dense matter in the region of non-perturbative QCD. In HICs, two asymmetric nuclei are collided with each other and for a short interval of time, a fireball containing quark gluon plasma (QGP) comes into existence. Within a very short interval, the fireball expands and  converts into an ensemble of particles which consists of nucleons, hyperons, and mesons,  collectively  known as hadronic matter \cite{Kumar2019}. In a hadronic medium, the baryons and mesons are the degrees of freedom therefore only non-perturbative physics can be applied here. Furthermore, due to the presence of the strange particles in the medium, it is very important to include the strangeness fraction while studying the properties of mesons in the hadronic matter. We see significant progress to understand the properties of strange dense matter at a moderate temperature with the construction of future experiments such as CBM at Facility for Antiproton and Ion Research (FAIR), Nuclotron-based Ion Collider Facility (NICA) at Dubna, Russia and, J-PARC in Japan \cite{Kumar2019,Rapp2010}. 

The in-medium study of light vector mesons ($\rho$, $\omega$ and $\phi$) is of interest to theoretical and experimental researchers \cite{Kim2020,Mishra2015,Mishra2019a,Shivam2019,Leupold2010,Hayano2010,
Krein2016,Martinez2016,Ko1992,Li1995} because of their role in exploring the dilepton production in HICs. The dilepton production  is considered as promising observable because of its weak interaction with baryons and mesons \cite{Xiong1990,Xiong1990a,Ko1992,Korpa1990,Ko1989,Gale1987,Xia1988}. Among these mesons, there is a particular interest in $\phi$ meson,   because of   its strong interaction with nucleons and $u$/$d$ quarks although it has pure strange content ($s\bar s$) \cite{Martinez2016}. Due to its strange nature, it is also imperative to study its interactions with strange  baryons and mesons. In literature, using QCD Van der Waals forces, the $\phi$ meson  is used to test the multi gluon exchange theory \cite{Sibirstev2006} and it has implications to understand the dark matter \cite{Gubler2014,Bottino2002,Ellis2008}
as well, which is beyond  the regime of  QCD physics. There is also a possibility of $\phi$-mesic nuclei formation due to its negative mass-shift in nuclear matter \cite{Martinez2017,Martinez2016,Buhler2010,Ohnishi2014,Jlab}.  The strong interaction of $\phi$ with $u$/$d$ quarks occurs due to the interplay with $K\bar K$ pairs, therefore, the in-medium properties of the kaon and antikaon play a crucial role. It was Kaplan and Nelson, who initiated the study of in-medium properties of kaons and antikaons \cite{Kaplan1986}.
They observed a negative mass shift of antikaons in the neutron star medium and suggested the possibility of antikaon condensation.   The downward mass-shift comes from strong
attractive interaction of antikaons with nucleons. Due to this attractive interaction, the effective energy of the mesons becomes low  by  transforming in the attractive scalar field. The annihilation of $K \bar K$
into a dilepton mainly yields through the $\phi$ meson, therefore dilepton production in HICs helps us to understand   the properties of the $\phi$ meson in the dense hadronic
medium \cite{Xiong1990,Xiong1990a,Ko1992,Korpa1990,Ko1989,Gale1987,Xia1988}. Due to the relevance of $K$ and $\bar K$ isospin doublets in   heavy-ion collisions, a lot of theoretical \cite{Mishra2009,Li1997,Ko2001,Pal2001,Cassing1997,
Bratkovskaya1997,Cassing1999,Lutz1998,Lutz2002,Lutz2002a,Ramos2000,Tolos2002} and experimental \cite{Laue1999,Menzel2000,Sturm2001,Forster2002} investigations have been made.
The free space antikaon-nucleon scattering amplitudes are obtained  from a covariant   unitarized chiral coupled-channel approaches  which
include the method of partial waves  systematically \cite{Lutz1998,Ramos2000,Tolos2001,Tolos2006,Lutz2008}. The  evaluation
of the kaon self-energy from this mechanism has successfully described
the $K^-$ meson interaction in the hadronic matter.  At 
nuclear saturation density, an attractive potential of about 40 to 60 MeV is obtained from these calculations.  Using the chiral SU(3) model, the properties of kaons and antikaons are studied in isospin asymmetric nuclear matter at finite temperature in Ref. \cite{Mishra2008}, and in strange matter at zero temperature  \cite{Mishra2009}. In these articles, the in-medium self energies  of $K$ and $ \bar K$ mesons are studied and  an attractive 
in-medium optical potential is found.
 
 The in-medium theoretical observations of  $\phi$ meson mass and decay width  has been studied extensively in the literature. Several authors have speculated  a small downward mass shift and broadening in the decay width of $\phi$ meson \cite{Ko1992,Klingl1998,Hatsuda1991,Hatsuda1996,Oset2000,Cabrera2002,Martinez2016}. By considering the contributions of the kaon-antikaon loop to the self-energy,
Ko $et.al.$~\cite{Ko1992} used chiral perturbation theory to calculate the density-dependent kaon mass  and found that at nuclear  saturation
density, $\rho_0$,
the $\phi$ meson mass decreases very little (at most $2\%$), and the width 
has the value $\approx 25$~MeV. They also observed that for large densities  the decay width  broadens substantially.  In Ref.~\cite{Klingl1998}, at $\rho_0$, Klingl $et.al.$ reports a 
downward mass shift in $\phi$ mass ( $ < 1\%$) and a broadened decay width of 45 MeV. Using the QCD sum rule approach with linear density approximation,  
Hatsuda and Lee computed the in-medium 
$\phi$ mass, 
and predicted a small decrease  at nuclear saturation density ~\cite{Hatsuda1991,Hatsuda1996}.
The large broadening of $\phi$ decay width is also predicted by other investigators as in
Ref.~\cite{Oset2000}, Oset $et.al.$ predicted a decay width of 22 MeV and in Ref.~\cite{Cabrera2002}  a decay width of 30 MeV was observed.
More recently, in Ref.~\cite{Martinez2016}, Martinez $et.al.$ reported a downward mass shift of
$25$ MeV and a large broadening width of 32.8 MeV for a cut-off parameter of 3000 MeV at nuclear saturation density. In most of the experiments, 
a large broadening of the in-medium decay width has been investigated \cite{Muto2005,Ishikawa2004,Mibe2007,Qian2009}. At nuclear saturation density, the KEK-E325 collaboration computed a decrement in mass  ($3.4\%$) and  increase in the in-medium decay width  ($\approx 14.5$ MeV)  of $\phi$ meson ~\cite{Muto2005}.
Whereas in Ref.~\cite{Ishikawa2004}  SPring8 reported  a large value of $\phi N$ cross-section in the medium which results in a decay
width of 35 MeV, which is in close  agreement with the experiments \cite{Mibe2007,Qian2009}. To have a more clear picture, further experimental efforts are needed to understand $\phi$ meson in the medium.

In the present article, we report the results of the in-medium  $\phi$  meson mass and decay width in the hot asymmetric strange hadronic matter by taking into account the medium induced  kaon and antikaon masses. The in-medium $K$ and $\bar K$ properties are incorporated by a chiral effective Lagrangian using the chiral SU(3) model \cite{Mishra2009,Kumar2020}. We  calculate the in-medium masses of $K$ and $\bar K$ mesons at finite temperatures in the strange hadronic matter and  use those as input to calculate the in-medium mass and decay width of $\phi$ meson. The chiral model is a non-perturbative hadron based model to describe the in-medium properties of hadronic matter \cite{Papazoglou1999,Kumar2014,Kumar2019a,Mishra2019,
Mishra2004a,Mishra2004,Mishra2006,Mishra2008,
Kumar2020a,Reddy2018,Dhale2018,Kumar2015,Chhabra2017,Chhabra2018,Kumar2010}. It has also been applied to study the effect of the magnetic field on the in-medium properties of  quarkonia \cite{Kumar2019,Kumar2019a} and open charm mesons \cite{Kumar2020,Kumar2020a}. To study the $K \bar K$ loop contributions in the $\phi$ meson decay, we use effective Lagrangian of $\phi K \bar K$ interactions and solve the loop integral by using regularization techniques \cite{Martinez2016,Krein2010}. In the current work, we include the contributions from $K$ and $\bar K$ loop by utilizing the in-medium masses $m_{K^+}^{*}$, $m_{K^0}^{*}$, $m_{K^-}^{*}$ and $m_{\bar K^0}^{*}$  which is different from previous works as in Ref. \cite{Martinez2016,Martinez2017}, the $\bar K$ contribution to the $K \bar K$ loop   was suppressed by equalising the mass via relation $m^*_K$=$m^*_{\bar K}$.

 The layout of the present paper is as follows: In the next subsection \ref{subsec2.1}, we will concisely discuss the  methodology to obtain the in-medium scalar and vector fields in the hyperonic matter.  In the  subsection \ref{subsec2.2}, we will calculate the induced mass of kaon and antikaon via interactions of chiral model fields. The theoretical approach to calculate the in-medium mass and decay width of $\phi$ meson will be discussed in subsection \ref{subsec2.3}. In the forthcoming section  \ref{sec:3}, quantitative results of the present findings will be discussed and in closing, we will conclude our work in section \ref{sec:4}.

\section{ Methodology }

We use chiral SU(3) model to study the impact of isospin asymmetry and strangeness fraction on the scalar and vector fields, which is further used to calculate the in-medium mass of kaon and antikaon. Moreover, the $\phi$ meson mass and decay width is calculated from self consistent Lagrangian approach. In the forthcoming subsections, we briefly narrate the formalism to obtain the results. 
\subsection{THE HADRONIC CHIRAL SU(3) MODEL}
\label{subsec2.1}
The  hadronic chiral effective Lagrangian is given as
\be
{\cal L}_{chiral} = {\cal L}_{kin} + \sum_{ M =S,V} {\cal L}_{BM}
          + {\cal L}_{vec} + {\cal L}_0 + {\cal L}_{SB}.
\label{genlag} \ee

 As a description of hadronic matter, the model comprises the fundamental QCD features such as  trace anomaly  and non-linear realization of chiral symmetry \cite{Weinberg1968,Coleman1969,Zschiesche1997,Bardeen1969,
Kumar2020,Papazoglou1999,Kumar2019}.  In this model, the isospin asymmetry of the matter is introduced  by the inclusion of the scalar isovector field $\delta$ and vector-isovector field $\rho$ \cite{Kumar2020} and the impact of strangeness is measured by incorporating the scalar field $\zeta$ and  vector field $\phi$. The broken scale invariance property of QCD  is conserved by the insertion of the scalar dilaton field $\chi$ \cite{Papazoglou1999,Kumar2020}. To simplify, the effect of   fluctuations near phase transitions are neglected  by using mean-field approximation \cite{Kumar2020,Reddy2018}.  This model has been used successfully to
study the nuclear matter, hypernuclei, finite nuclei, and neutron
stars \cite{Weinberg1968,Coleman1969,Zschiesche1997,Bardeen1969,
Kumar2020,Papazoglou1999,Kumar2019}.

In the Eq.(\ref{genlag}), ${\cal L}_{kin}$ denotes the kinetic energy term, ${\cal L}_{BM}$ is the baryon-meson interaction term, 
 where $S$ and $V$ represents the scalar and vector mesons, respectively.  The term $ {\cal L}_{vec}$  produces the  vector meson mass through the interactions with scalar mesons and contains the quartic self-interaction terms, $ {\cal L}_{0}$ defines  the spontaneous chiral symmetry breaking, and  ${\cal L}_{SB} $ defines the explicit chiral symmetry breaking. There
exist the $D$-type (symmetric) and $F$-type (antisymmetric)  couplings for the baryon-vector meson interactions terms. Here we use the antisymmetric coupling \cite{Mishra2009,Kumar2015} as per the vector meson dominance model the $D$-type coupling should be less which also follows  the universality principle  \cite{Sakurai1969}. 
Moreover, we choose the medium parameters \cite{Mishra2009} so as to
dissociate nucleons from the strange  field $
\phi_\mu\sim\bar{s} \gamma_\mu s $, which leads to an ideal mixing between $\phi$ and $\omega$. 

By using Euler Lagrange equations for $\sigma$, $\zeta$, $\delta$,  $\omega$, $\rho$, $\phi$ and $\chi$ mesonic fields of the chiral model, we obtained the following  coupled equations of motions:

\begin{eqnarray}
 k_{0}\chi^{2}\sigma-4k_{1}\left( \sigma^{2}+\zeta^{2}
+\delta^{2}\right)\sigma-2k_{2}\left( \sigma^{3}+3\sigma\delta^{2}\right)
-2k_{3}\chi\sigma\zeta \nonumber\\
-\frac{d}{3} \chi^{4} \bigg (\frac{2\sigma}{\sigma^{2}-\delta^{2}}\bigg )
+\left( \frac{\chi}{\chi_{0}}\right) ^{2}m_{\pi}^{2}f_{\pi}
=\sum g_{\sigma i}\rho_{i}^{s} ,
\label{sigma}
\end{eqnarray}
\begin{eqnarray}
 k_{0}\chi^{2}\zeta-4k_{1}\left( \sigma^{2}+\zeta^{2}+\delta^{2}\right)
\zeta-4k_{2}\zeta^{3}-k_{3}\chi\left( \sigma^{2}-\delta^{2}\right)\nonumber\\
-\frac{d}{3}\frac{\chi^{4}}{\zeta}+\left(\frac{\chi}{\chi_{0}} \right)
^{2}\left[ \sqrt{2}m_{K}^{2}f_{K}-\frac{1}{\sqrt{2}} m_{\pi}^{2}f_{\pi}\right]
 =\sum g_{\zeta i}\rho_{i}^{s} ,
\label{zeta}
\end{eqnarray}
\begin{eqnarray}
k_{0}\chi^{2}\delta-4k_{1}\left( \sigma^{2}+\zeta^{2}+\delta^{2}\right)
\delta-2k_{2}\left( \delta^{3}+3\sigma^{2}\delta\right) +2k_{3}\chi\delta
\zeta \nonumber\\
 +   \frac{2}{3} d \chi^4 \left( \frac{\delta}{\sigma^{2}-\delta^{2}}\right)
=\sum g_{\delta i}\tau_3\rho_{i}^{s}  ,
\label{delta}
\end{eqnarray}

\begin{eqnarray}
\left (\frac{\chi}{\chi_{0}}\right) ^{2}m_{\omega}^{2}\omega+g_{4}\left(4{\omega}^{3}+12{\rho}^2{\omega}\right) =\sum g_{\omega i}\rho_{i}^{v}  ,
\label{omega}
\end{eqnarray}

\begin{eqnarray}
\left (\frac{\chi}{\chi_{0}}\right) ^{2}m_{\rho}^{2}\rho+g_{4}\left(4{\rho}^{3}+12{\omega}^2{\rho}\right)=\sum g_{\rho i}\tau_3\rho_{i}^{v}  ,
\label{rho}
\end{eqnarray}

  \begin{eqnarray}
\left (\frac{\chi}{\chi_{0}}\right) ^{2}m_\phi^2\phi+8g_4\phi^3&=&
\sum g_{\phi i}\rho_{i}^{v},
 \label{phi}  
\end{eqnarray}

and

\begin{eqnarray}
k_{0}\chi \left( \sigma^{2}+\zeta^{2}+\delta^{2}\right)-k_{3}
\left( \sigma^{2}-\delta^{2}\right)\zeta + \chi^{3}\left[1
+{\rm {ln}}\left( \frac{\chi^{4}}{\chi_{0}^{4}}\right)  \right]
+(4k_{4}-d)\chi^{3}
\nonumber\\
-\frac{4}{3} d \chi^{3} {\rm {ln}} \Bigg ( \bigg (\frac{\left( \sigma^{2}
-\delta^{2}\right) \zeta}{\sigma_{0}^{2}\zeta_{0}} \bigg )
\bigg (\frac{\chi}{\chi_0}\bigg)^3 \Bigg )+
\frac{2\chi}{\chi_{0}^{2}}\left[ m_{\pi}^{2}
f_{\pi}\sigma +\left(\sqrt{2}m_{K}^{2}f_{K}-\frac{1}{\sqrt{2}}
m_{\pi}^{2}f_{\pi} \right) \zeta\right] \nonumber\\
-\frac{\chi}{{{\chi_0}^2}}(m_{\omega}^{2} \omega^2+m_{\rho}^{2}\rho^2)  = 0.
\label{chi}
\end{eqnarray}

In above equations, the model parameters $k_i (i=1$ to $4)$ are fitted to regenerate the vacuum values of scalar fields \cite{Kumar2010} and $m_\pi$, $m_K$, $f_\pi$ and $f_K$  denote the masses and decay constants of pions and kaons, respectively. The values of these parameters along with other coupling constants which are fitted in the model to reproduce vacuum masses of baryon octet are tabulated in \cref{ccc}.  In these equations of motion, $\rho^{v}_{i}$ and $\rho^{s}_{i}$ characterize the  vector and scalar densities of $i^{th}$ baryons ($i=p,n, \Lambda, \Sigma ^\pm,
\Sigma ^0, \Xi ^-, \Xi ^0$) \cite{Kumar2019,Mishra2009} and are defined through relations:

\begin{eqnarray}
\rho_{i}^{v} = \gamma_{i}\int\frac{d^{3}k}{(2\pi)^{3}}  
\Bigg(\frac{1}{1+\exp\left[\beta(E^{\ast}_i(k) 
-\mu^{*}_{i}) \right]}-\frac{1}{1+\exp\left[\beta(E^{\ast}_i(k)
+\mu^{*}_{i}) \right]}
\Bigg),
\label{rhov0}
\end{eqnarray}

 and

\begin{eqnarray}
\rho_{i}^{s} = \gamma_{i}\int\frac{d^{3}k}{(2\pi)^{3}} 
\frac{m_{i}^{*}}{E^{\ast}_i(k)} \Bigg(\frac{1}{1+\exp\left[\beta(E^{\ast}_i(k) 
-\mu^{*}_{i}) \right]}+\frac{1}{1+\exp\left[\beta(E^{\ast}_i(k)
+\mu^{*}_{i}) \right]}
\Bigg),
\label{rhos0}
\end{eqnarray}
respectively, where $\beta = \frac{1}{kT}$ and $\gamma_i$ is the degeneracy factor. Through the temperature dependence in above relations, we introduce the 
temperature dependence in the values of scalar and vector fields and hence, masses of kaons and antikaons
and further, in the masses and decay width of $\phi$ mesons.
 In addition, in this model  the  isospin asymmetry and strangeness  is incorporated  through the parameters, $\eta = -\frac{\Sigma_i \tau_{3i} \rho^{v}_{i}}{2\rho_{B}}$ and $f_s = \frac{\Sigma_i \vert s_{i} \vert \rho^{v}_{i}}{\rho_{B}}$ respectively. Where  $\tau_3$, $\vert s_{i} \vert$ and $\rho_B$ symbolize the isospin quantum number (3$^{rd}$ component), number of strange quarks  and the total baryonic density respectively.

\begin{table}
\begin{tabular}{|c|c|c|c|c|c|c|c|c|c|}

\hline
$g_{\sigma n(p)}$  & $g_{\zeta n(p) }$  &  $g_{\delta n(p) }$  &
$g_{\omega n(p)}$ & $g_{\rho n(p)}$ & $g_{\sigma \Lambda}$ &$g_{\zeta \Lambda}$&  $g_{\delta \Lambda}$& 
$g_{\sigma \Sigma}$&  $g_{\zeta \Sigma}$\\

\hline 
10.56 & -0.46 & 2.48 & 13.35 & 5.48 &7.52&5.8&0&6.13&5.8 \\

\hline
$g_{\delta \Sigma}$  &  $g_{\delta \Sigma^0}$ &
$g_{\sigma \Xi}$ & $g_{\zeta \Xi}$ & $g_{\delta \Xi}$&$g_{\omega \Lambda}$&$g_{\omega \Sigma}$&$g_{\rho \Sigma}$&$g_{\rho \Sigma^0}$&$g_{\omega \Xi}$\\

\hline 
6.79 &0&3.78&9.14 &2.36 &$\frac{2}{3}$ $g_{\omega N}$&$\frac{2}{3}$ $g_{\omega N}$&$\frac{2}{3}$ $g_{\omega N}$&0& $\frac{1}{3}$ $g_{\omega N}$   \\ 

\hline
$g_{\rho \Lambda}$  &  $g_{\rho \Xi}$ &
$g_{\phi \Lambda}$ & $g_{\phi \Sigma}$ & $g_{\phi \Xi}$&$\sigma_0$ (MeV)& $\zeta_0$(MeV) & $\chi_0$(MeV)  & $d$ & $\rho_0$ ($\text{fm}^{-3}$)  \\

\hline 
0 &$\frac{1}{3}$ $g_{\omega N}$&$\frac{1}{3}$ $g_{\omega N}$&-$\frac{\sqrt 2}{3}$ $g_{\omega N}$&-$\frac{2 \sqrt 2}{3}$ $g_{\omega N}$ &-93.29 & -106.8 & 409.8 & 0.064 & 0.15  \\

\hline 
$m_\pi $(MeV) &$ m_K$ (MeV)&$ f_\pi$(MeV)  & $f_K$(MeV) & $g_4$&$k_0$ & $k_1$ & $k_2$ & $k_3$ & $k_4$    \\ 
\hline 
139 & 494 & 93.29 & 122.14 & 79.91 &2.53 & 1.35 & -4.77 & -2.77 & -0.218   \\

\hline
$m_\omega$ (MeV) & $m_\rho$ (MeV) & $m_\phi$ (MeV) &  &  &&&&& \\ 
\hline 
783 & 783 & 1020 & & &&&&& \\ 
\hline

\end{tabular}
\caption{Various parameters used in the present calculations in strange hadronic matter \cite{Kumar2011}.} \label{ccc}
\end{table} 

\subsection{KAON AND ANTIKAON INTERACTIONS IN THE CHIRAL MODEL}
\label{subsec2.2}

In this subsection, we evaluate the in-medium mass of  $K (\bar K)$ via dispersion relation \cite{Mao1999} 
in hot asymmetric strange  hadronic matter  \cite{Mishra2006,Kumar2015}.  As we discussed earlier,  
the in-medium masses of kaons and antikaons within chiral SU(3) model are studied in asymmetric nuclear matter at finite temperature whereas in strange matter at zero temperature only.  Whereas in the present work we will study these properties in
strange matter at finite temperature. 
 The scalar and vector fields modify the   scalar and vector densities of the baryons which further modifies the self-energy of the kaons and antikaons.  

The interaction Lagrangian density for kaons and antikaons can be written as \cite{Mishra2009}
\begin{eqnarray}
\cal L _{KB} & = & -\frac {i}{4 f_K^2} \Big [\Big ( 2 \bar p \gamma^\mu p
+\bar n \gamma ^\mu n -\bar {\Sigma^-}\gamma ^\mu \Sigma ^-
+\bar {\Sigma^+}\gamma ^\mu \Sigma ^+
- 2\bar {\Xi^-}\gamma ^\mu \Xi ^-
- \bar {\Xi^0}\gamma ^\mu \Xi^0 \Big)
\nonumber \\
& \times &
\Big(K^- (\partial_\mu K^+) - (\partial_\mu {K^-})  K^+ \Big )
\nonumber \\
& + &
\Big ( \bar p \gamma^\mu p
+ 2\bar n \gamma ^\mu n +\bar {\Sigma^-}\gamma ^\mu \Sigma ^-
-\bar {\Sigma^+}\gamma ^\mu \Sigma ^+
- \bar {\Xi^-}\gamma ^\mu \Xi ^-
- 2 \bar {\Xi^0}\gamma ^\mu \Xi^0 \Big)
\nonumber \\
& \times &
\Big(\bar {K^0} (\partial_\mu K^0) - (\partial_\mu {\bar {K^0}})  K^0 \Big )
\Big ]
\nonumber \\
 &+ & \frac{m_K^2}{2f_K} \Big [ (\sigma +\sqrt 2 \zeta+\delta)(K^+ K^-)
 + (\sigma +\sqrt 2 \zeta-\delta)(K^0 \bar { K^0})
\Big ] \nonumber \\
& - & \frac {1}{f_K}\Big [ (\sigma +\sqrt 2 \zeta +\delta)
(\partial _\mu {K^+})(\partial ^\mu {K^-})
+(\sigma +\sqrt 2 \zeta -\delta)
(\partial _\mu {K^0})(\partial ^\mu \bar {K^0})
\Big ]
\nonumber \\
&+ & \frac {d_1}{2 f_K^2}(\bar p p +\bar n n +\bar {\Lambda^0}{\Lambda^0}
+\bar {\Sigma ^+}{\Sigma ^+}
+\bar {\Sigma ^0}{\Sigma ^0}
+\bar {\Sigma ^-}{\Sigma ^-}
+\bar {\Xi ^-}{\Xi ^-}
+\bar {\Xi ^0}{\Xi ^0}
 )\nonumber \\
&\times & \big ( (\partial _\mu {K^+})(\partial ^\mu {K^-})
+(\partial _\mu {K^0})(\partial ^\mu {\bar {K^0}})
\big )
\nonumber \\
&+& \frac {d_2}{2 f_K^2} \Big [
(\bar p p+\frac {5}{6} \bar {\Lambda^0}{\Lambda^0}
+\frac {1}{2} \bar {\Sigma^0}{\Sigma^0}
+\bar {\Sigma^+}{\Sigma^+}
+\bar {\Xi^-}{\Xi^-}
+\bar {\Xi^0}{\Xi^0}
) (\partial_\mu K^+)(\partial^\mu K^-) 
\nonumber \\
 &+ &(\bar n n
+\frac {5}{6} \bar {\Lambda^0}{\Lambda^0}
+\frac {1}{2} \bar {\Sigma^0}{\Sigma^0}
+\bar {\Sigma^-}{\Sigma^-}
+\bar {\Xi^-}{\Xi^-}
+\bar {\Xi^0}{\Xi^0}
) (\partial_\mu K^0)(\partial^\mu {\bar {K^0}})
\Big ],
\label{lagd}
\end{eqnarray}
with  the vectorial interaction term
(Weinberg-Tomozawa term) as first term which is obtained from the kinetic part of the interaction Lagrangian. 
 The
 second term is obtained from explicit symmetry breaking 
and the third term is acquired by the  kinetic terms of pseudoscalar meson of the chiral effective Lagrangian \cite{Mishra2006,Mishra2008}. The fourth and fifth terms 
are called range terms which basically arise from the baryon meson interaction Lagrangian of chiral model  \cite{Mishra2004a,Mishra2006} and are given as
\begin{equation}
{\cal L }_{d_1}^{BM} =\frac {d_1}{2} Tr (u_\mu u ^\mu)Tr( \bar B B),
\end{equation}
and,
\begin{equation}
{\cal L }_{d_2}^{BM} =d_2 Tr (\bar B u_\mu u ^\mu B),
\end{equation}
where $B$ denotes the baryon octet.
The dispersion relation for kaon and antikaon is obtained  by Fourier transformation of interaction Lagrangian and given by
\begin{equation}
-\omega^2+ {\vec k}^2 + m_{K (\bar K)}^2 -\Pi^*(\omega, |\vec k|)=0,
\label{drk}
\end{equation}

where $\Pi^*$ symbolize the in-medium self-energy of  kaon and antikaon, and  for the kaon isospin doublet,
($K^+$,$K^0$), 
it is explicitly given as
\begin{eqnarray}
\Pi^*_K (\omega, |\vec k|) &= & -\frac {1}{4 f_K^2}\Big [3 (\rho^v_p +\rho^v_n)
\pm (\rho^v_p -\rho^v_n) \pm 2 (\rho^v_{\Sigma^+}-\rho^v_{\Sigma^-})
-\big ( 3 (\rho^v_{\Xi^-} +\rho^v_{\Xi^0}) \pm (\rho^v_{\Xi^-} -\rho^v_{\Xi^0})
\big)
\Big ] \omega\nonumber \\
&+&\frac {m_K^2}{2 f_K} (\sigma ' +\sqrt 2 \zeta ' \pm \delta ')
\nonumber \\ & +& \Big [- \frac {1}{f_K}
(\sigma ' +\sqrt 2 \zeta ' \pm \delta ')
+\frac {d_1}{2 f_K ^2} (\rho_s ^p +\rho_s ^n
+{\rho^s} _{\Lambda^0}+{\rho^s} _{\Sigma^+}+{\rho^s} _{\Sigma^0}
+{\rho^s} _{\Sigma^-} +{\rho^s} _{\Xi^-} +{\rho^s} _{\Xi^0}
)\nonumber \\
&+&\frac {d_2}{4 f_K ^2} \Big (({\rho^s} _p +{\rho^s} _n)
\pm   ({\rho^s} _p -{\rho^s} _n)
+{\rho^s} _{\Sigma ^0}+\frac {5}{3} {\rho^s} _{\Lambda^0}
+ ({\rho^s} _{\Sigma ^+}+{\rho^s} _{\Sigma ^-})
\pm ({\rho^s} _{\Sigma ^+}-{\rho^s} _{\Sigma ^-})\nonumber \\
 &+ & 2 {\rho^s} _ {\Xi^-}+
2 {\rho^s} _ {\Xi^0}
\Big )
\Big ]
(\omega ^2 - {\vec k}^2),
\label{sek}
\end{eqnarray}
where the $\pm$ signs mention the self-energy for $K^+$ and $K^0$ respectively.  In the above expression  the fluctuations $\sigma'(=\sigma-\sigma _0)$,
$\zeta'(=\zeta-\zeta_0)$ and  $\delta'(=\delta-\delta_0)$ indicate the  digression of the field expectation values 
from their vacuum expectation values. Also,  $m_{K(\bar K)}$ in Eq.(\ref{drk}) denotes the vacuum mass of kaon (antikaon).

In a similar manner, the in-medium self-energy for antikaon isospin doublet, ($K^-$,$\bar {K^0}$),
is evaluated as
\begin{eqnarray}
\Pi^*_{\bar K} (\omega, |\vec k|) &= & \frac {1}{4 f_K^2}\Big [3 (\rho^v_p +\rho^v_n)
\pm (\rho^v_p -\rho^v_n) \pm 2 (\rho^v_{\Sigma^+}-\rho^v_{\Sigma^-})
- \big ( 3 (\rho^v_{\Xi^-} +\rho^v_{\Xi^0}) \pm (\rho^v_{\Xi^-} -\rho^v_{\Xi^0})
\big)
\Big ] \omega\nonumber \\
&+&\frac {m_{\bar K}^2}{2 f_{\bar K}} (\sigma ' +\sqrt 2 \zeta ' \pm \delta ')
\nonumber \\ & +& \Big [- \frac {1}{f_{\bar K}}
(\sigma ' +\sqrt 2 \zeta ' \pm \delta ')
+\frac {d_1}{2 f_{\bar K}^2} (\rho_s ^p +\rho_s ^n
+{\rho^s} _{\Lambda^0}+{\rho^s} _{\Sigma^+}+{\rho^s} _{\Sigma^0}
+{\rho^s} _{\Sigma^-} +{\rho^s} _{\Xi^-} +{\rho^s} _{\Xi^0}
)\nonumber \\
&+&\frac {d_2}{4 f_K ^2} \Big (({\rho^s} _p +{\rho^s} _n)
\pm   ({\rho^s} _p -{\rho^s} _n)
+{\rho^s} _{\Sigma ^0}+\frac {5}{3} {\rho^s} _{\Lambda^0}
+ ({\rho^s} _{\Sigma ^+}+{\rho^s} _{\Sigma ^-})
\pm ({\rho^s} _{\Sigma ^+}-{\rho^s} _{\Sigma ^-})\nonumber \\
 &+ & 2 {\rho^s} _ {\Xi^-}+ 2 {\rho^s} _ {\Xi^0}
\Big )
\Big ]
(\omega ^2 - {\vec k}^2),
\label{sekb}
\end{eqnarray}
with the $\pm$ signs for $K^-$ and $\bar {K^0}$ respectively.

In strange hadronic  matter, the in-medium mass of $K (\bar K)$ meson is evaluated by solving the Eq. (\ref{drk}) under the condition,
$m_{K(\bar K)}^*=\omega(|\vec k|$=0). The parameters $d_1$ and $d_2$ in the  expression of self energies
are taken as $ 2.56/m_K $ and $ 0.73/m_K $ 
respectively \cite{Mishra2019}, 
 fitted with  the help of experimental values  of
 kaon-nucleon ($KN$) scattering length \cite{Barnes1994}. In the present work,  the vacuum value of $K^+(K^-)$  mass is taken as 494 MeV whereas for $K^0(\bar K^0)$ it is taken as 498 MeV.
\subsection{\label{subsec2.3} IN-MEDIUM MASS AND DECAY WIDTH OF $\phi$ MESON }

 \begin{figure}[h]
\includegraphics[scale=0.1]{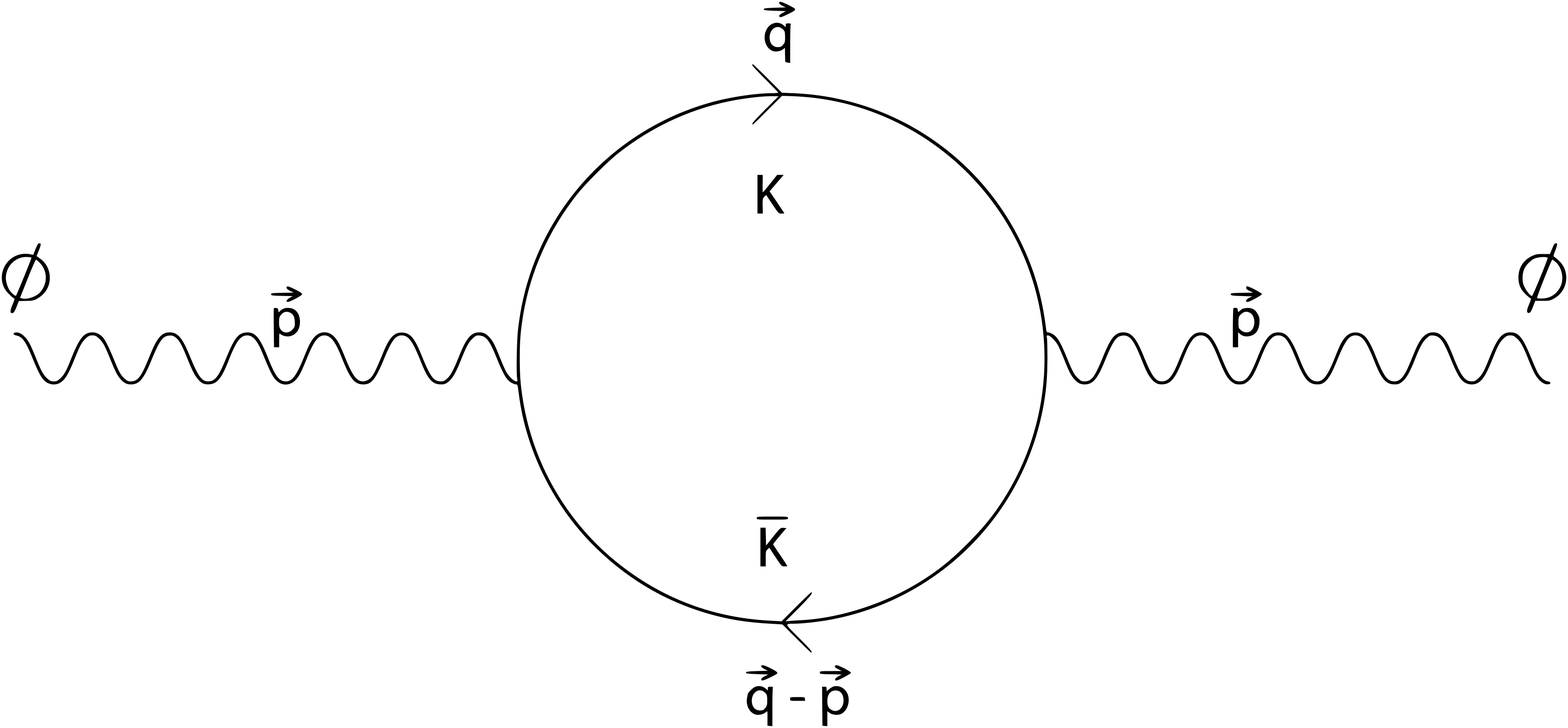}
\caption{ $\phi K \bar{K}$ interaction at one loop level.}
\label{loop}
\end{figure}

In this subsection, we first compute the $\phi$ meson in-medium self-energy for the decay process $\phi$ $\rightarrow$ $K \bar K$ at one loop level (see \cref{loop}). The interaction Lagrangian $\mathcal{L}_{int}$ \cite{{Ko1992},{Klingl1996}}  is given as
\begin{equation}
 \label{eqn:Lint}
\mathcal{L}_{int} = \mathcal{L}_{\phi K \bar K},
\end{equation}
with

\begin{equation}
\label{eqn:phikk}
\mathcal{L}_{\phi K \bar {K}} = i g_{\phi}\phi^{\mu}
\left[ \bar K(\partial_{\mu} K)-(\partial_{\mu} \bar K)K\right].
\end{equation}

In above, $g_\phi$ is the coupling constant and $K \left(\begin{array}{c} K^{+} \\ K^{0} \end{array} \right),$ and $\bar K \left(K^{-}\;\overline{K}^{0}\;\right)$ are the isospin doublets of kaons and antikaons.
In our present work we haven't considered the interactions of type $\phi\phi K \bar{K}$
as these have very little contribution to in-medium masses and decay width as compared to $\phi K \bar K$ interactions \cite{Martinez2016}.

 In the rest frame of $\phi$ meson, the scalar part of the in-medium self energy for the loop diagram  can be written as
\begin{equation}
\label{eqn:phise}
i\Pi^*_{\phi}(p)=-\frac{8}{3}g_{\phi}^{2}\int \frac {d^4q}{(2\pi)^4} \vec{q}^{\,2}
D_{K}(q)D_{\bar K}(q-p) \, ,
\end{equation}
where $D_{K}(q)$=$\left(q^{2}-m_{K}^{*^{2}}+i\epsilon\right)^{-1}$ is the
kaon propagator and $D_{\bar K}(q$-$p)$=$\left((q-p)^{2}-m_{\bar K}^{*^{2}}+i\epsilon\right)^{-1}$ is the antikaon propagator;  $p=(p^{0}=m^*_{\phi},\vec{0})$ is 
the $\phi$ meson four-momentum vector, 
with $m^*_{\phi}$ denoting the in-medium $\phi$ meson mass; $m^*_{K}$(=$\frac{m_{K^+}^{*}+m_{K^0}^{*}}{2}$) and $m^*_{\bar K}$(=$\frac{m_{K^-}^{*}+m_{\bar K^0}^{*}}{2}$) are the masses of kaon and antikaon respectively. 
The values of $m_{K^+}^{*}$, $m_{K^0}^{*}$, $m_{K^-}^{*}$ and $m_{\bar K^0}^{*}$ will be solved using Eq.(\ref{drk}) and the in-medium mass of the $\phi$ is determined from the real part of $\Pi^*_{\phi}(p)$ by following relation \cite{Martinez2016}
\begin{equation}
\label{eqn:phimassvacuum}
m_{\phi}^{*^{2}}=\left(m_{\phi}^{0}\right)^{2}+\Re\Pi^*_{\phi}(m_{\phi}^{*^{2}}),
\end{equation}
where $m_{\phi}^{0}$ being the bare mass of the $\phi$ meson. The real part of self-energy can be written as \cite{Martinez2016}
\begin{equation}
\label{eqn:repiphi}
\Re\Pi^*_{\phi}=-\frac{4}{3}g_{\phi}^{2} \, \mathcal{P}\!\!
\int \frac {d^3q} {(2\pi)^3} \vec{q}^{\,2}\frac{(E^*_K+E^*_{\bar K})}{E^*_{K} E^*_{\bar K} ((E^*_K+E^*_{\bar K})^2-m_{\phi}^{*^2})} \, ,
\end{equation}
with $\mathcal{P}$ denotes the principal value  of the 
integral Eq.~(\ref{eqn:phise}), $E^*_{K}=(\vec{q}^{\,2}+m_{K}^{*^2})^{1/2}$ and $E^*_{\bar K}=(\vec{q}^{\,2}+m_{\bar K}^{*^2})^{1/2}$. 

%
%

The integral in Eq.(\ref{eqn:repiphi}) is divergent and to avoid the singularities we 
regularized the integral with the help of a phenomenological 
form factor with a cut-off parameter $\Lambda_{c}$ \cite{Krein2010}, whose value is taken as 3 GeV in the present investigation. The integral after regularization is given as

\begin{equation}
\label{eqn:regphi}
\Re\Pi^*_{\phi}=-\frac{4}{3}g_{\phi}^{2} \, \mathcal{P}\!\!
\int^{\Lambda_c}_{0}  \frac {d^3q} {(2\pi)^3} \vec{q}^{\,4}\left( \frac{\Lambda^2_c+m_{\phi}^{*^2}}{\Lambda^2_c+4E_{K}^{*^2}}\right)^4 \frac{(E^*_K+E^*_{\bar K})}{E^*_{K} E^*_{\bar K} ((E^*_K+E^*_{\bar K})^2-m_{\phi}^{*^2})} \, .
\end{equation}

 The value of coupling constant $g_{\phi}$ is determined as 4.539 from the empirical  
width of the $\phi$ meson in vacuum \cite{PDG2015}. The bare mass of $\phi$ is fixed through the constant $g_\phi$ and the vacuum mass of $\phi$ meson, which is taken as 1019.461 MeV \cite{PDG2015}. The decay width of the $\phi$ meson  is calculated from imaginary part of the self energy $\Im\Pi^*_{\phi}$, and is given in terms of $\phi$, $K$ and $\bar K$ mass \cite{Li1995}
\begin{equation}
\label{eqn:phidecaywidth}
\Gamma^*_{\phi}  = \frac{g_{\phi}^{2}}{24\pi
} \frac{1}{m_{\phi}^{*^5}} 
\left((m_{\phi}^{*^2}-(m_{K}^{*}+m_{\bar K}^{*})^2)(m_{\phi}^{*^2}-(m_{K}^{*}-m_{\bar K}^{*})^2)\right)^{3/2} \, .
\end{equation}

\section{Results and Discussions}
\label{sec:3}

  In this section, we will discuss the numerical results obtained in the present work. 
  First, we will discuss the in-medium dependence of scalar fields in subsection \ref{subsec:3.1}. In 
  subsection \ref{subsec:3.2} density, temperature , isospin asymmetry and strangeness fraction dependence of masses of kaons and antikaons will be presented and finally in subsection
  \ref{subsec:3.3}  will be devoted to present the results of $\phi$ meson masses and decay width.
 
  \subsection{Scalar Fields of the Chiral Model in Strange Hadronic Matter}
 \label{subsec:3.1}

\begin{figure}[h]
\includegraphics[width=16cm,height=21cm]{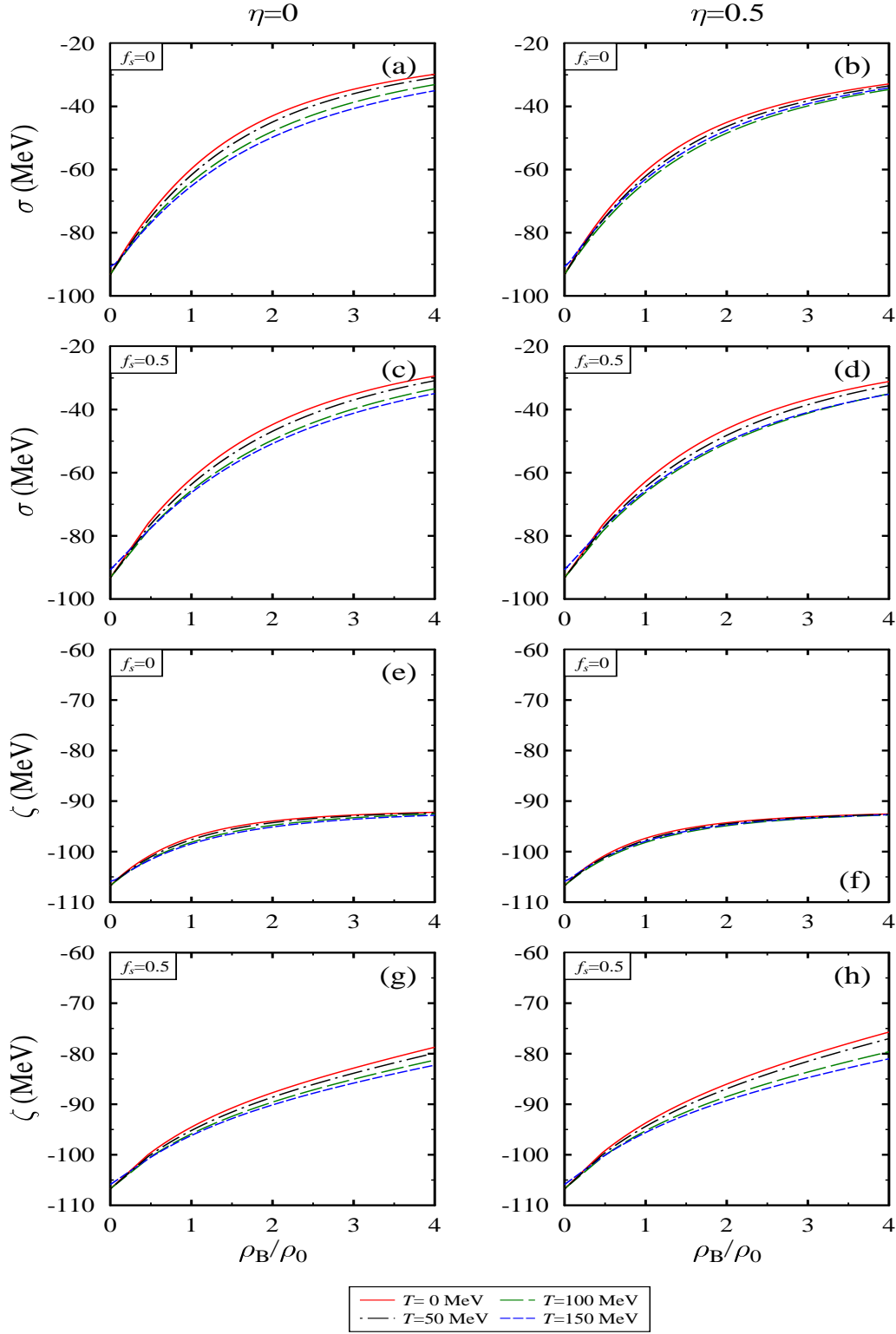}
\caption{(Color online) The in-medium $\sigma$ and $\zeta$ fields in nuclear and hyperonic matter. }
\label{fieldsT0}
\end{figure}

 As discussed in the subsection \ref{subsec2.1}, within the chiral model we have solved the coupled equations of motion of $\sigma$, $\zeta$, $\delta$, $\chi$, $\omega$, $\rho$ and $\phi$ mesonic fields. In \cref{fieldsT0}, we plot the variation of the scalar fields $\sigma$ and $\zeta$, as a function of baryonic density at finite values of temperature. We also observe the effect of strangeness fraction and isospin asymmetry in this plot. For all  combinations of $\eta$ and $f_s$, we see that the magnitude of $\sigma$ and $\zeta$ fields decrease linearly up to nuclear saturation density ($\rho_0$) and afterward it decreases slowly with further increase in baryonic density. However, the $\zeta$ field modifies very less as compared to $\sigma$ field, for example,  in a symmetric and non-strange medium at zero temperature and $\rho_B$=4$\rho_0$, the value of $\sigma$ field changes by 67$\%$ (as compared with its vacuum value) whereas $\zeta$ field modifies by 14$\%$ only. Furthermore, considering the effect of isospin asymmetry of the medium with respect to density, at a particular temperature, the scalar-isoscalar $\sigma$ field shows good $\eta$ dependence whereas strange scalar-isoscalar $\zeta$ field   shows negligible $\eta$ dependence. This is because of the quark content of the respective field, the former mesonic field contains $u$ and $d$ quark which interacts with the medium asymmetry, on the other hand, the latter contains strange quark pair ($s\bar s$). 
  This picture becomes opposite in the strange medium, as in the presence of hyperons, the strange $\zeta$ field shows appreciable modifications whereas   the non-strange $\sigma$ field shows very little variation. For example, in a symmetric and strange medium at zero temperature and $\rho_B$=4$\rho_0$, the value of $\zeta$ field changes by 16$\%$ (as compared with its value at $f_s$=0) whereas $\sigma$ field modifies by 2$\%$ only. In  symmetric  medium if we move from zero to non-zero temperature, we observe that in high density regime the value of $\sigma$ field modifies appreciably in both strange and non-strange medium, at a specific value of density, the magnitude of $\sigma$ field increase with the increase in temperature. Whereas for the non-strange medium, the strange $\zeta$ field shows negligible $T$ dependence but it turns out to be appreciable in the strange medium. This is because of the presence of strange content in the medium (hyperons) modifies the scalar densities of the hyperons (which depends upon the Fermi distribution functions \cite{Kumar2019}) therefore the strange scalar field $\zeta$. As discussed earlier, in the present investigation, the scalar and vector fields are calculated  from the coupled equations of motion  that contain the expressions of scalar and vector densities of baryons.
 
\begin{figure}[h]
\includegraphics[width=16cm,height=21cm]{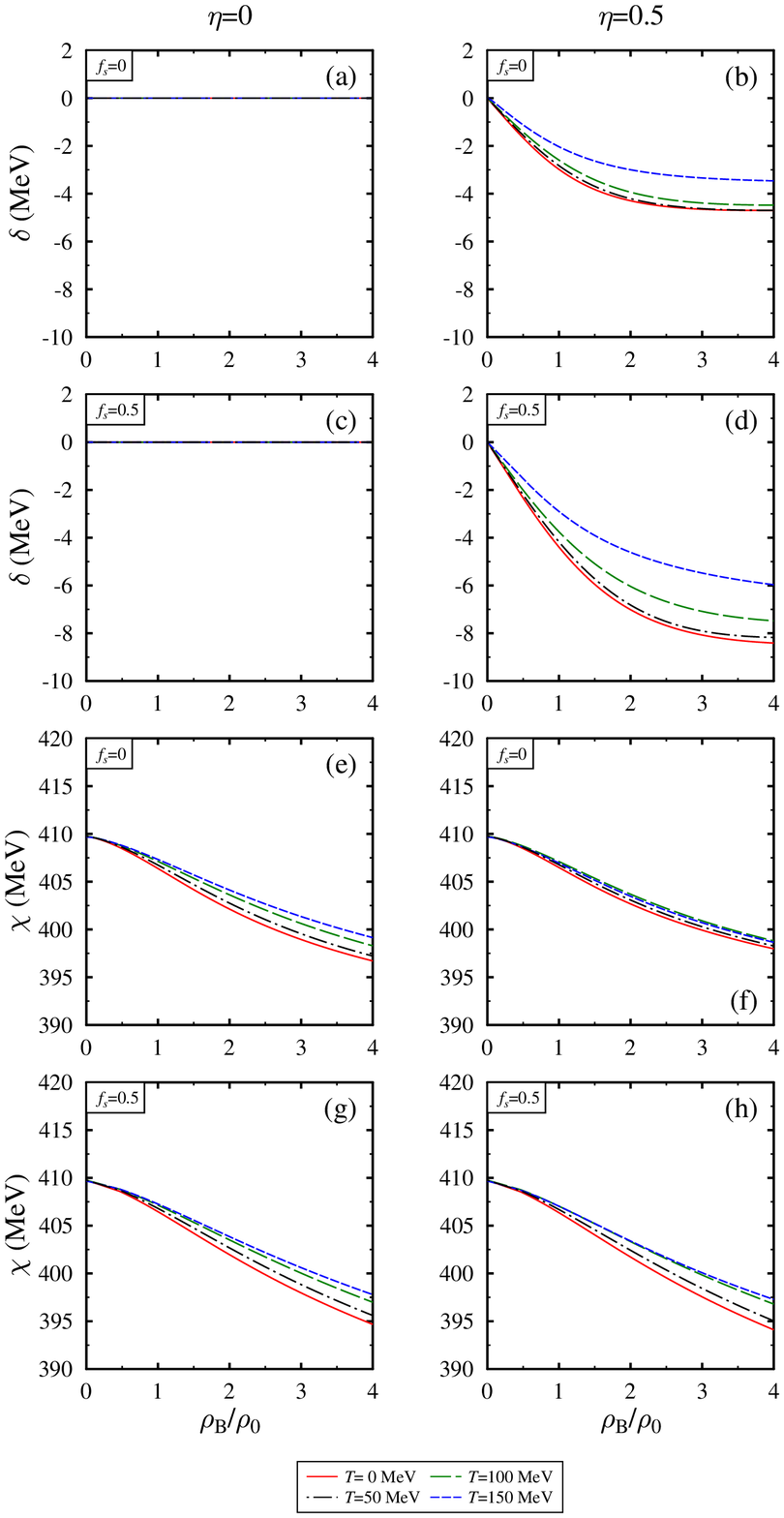}
\caption{(Color online) The in-medium $\delta$ and $\chi$ fields in nuclear and hyperonic matter. }
\label{fieldsT100}
\end{figure}

 
 In \cref{fieldsT100}, the variation of in-medium $\delta$ and $\chi$ fields are shown for the same medium attributes. As discussed earlier, the asymmetry dependence is introduced in this model by the incorporation of $\delta$ field and $\eta$ parameter, therefore the $\delta$ field shows appreciable variations in the asymmetric matter  but no modification  in the symmetric matter. At $\eta$=0.5 and non-strange medium, the magnitude of $\delta$ increases with density, and it pronounces more in the presence of strange baryons. For the finite baryonic density, the $\delta$ undergo less drop at high temperature  as compared to the $T$=0 situation. The dilaton  field $\chi$, which is introduced in the model to mimic the trace anomaly property of QCD \cite{Kumar2010}, varies less with the increase in baryonic density. The magnitude of $\chi$ field decreases as a function baryonic density and it shows variations due to asymmetry and strangeness in the high density regime. This is because the $\chi$ field is solved simultaneously in the coupled equations of motion along with other medium fields \cite{Kumar2020}. In the present  chiral SU(3)  model, the modifications of scalar fields with the medium's temperature are consistent with the results of
chiral quark mean-field model \cite{Wang2001}.
 
%
%

   \subsection{Kaons and Antikaons in Strange Matter at Finite Temperature}
 \label{subsec:3.2}

\begin{figure}
\includegraphics[width=16cm,height=21cm]{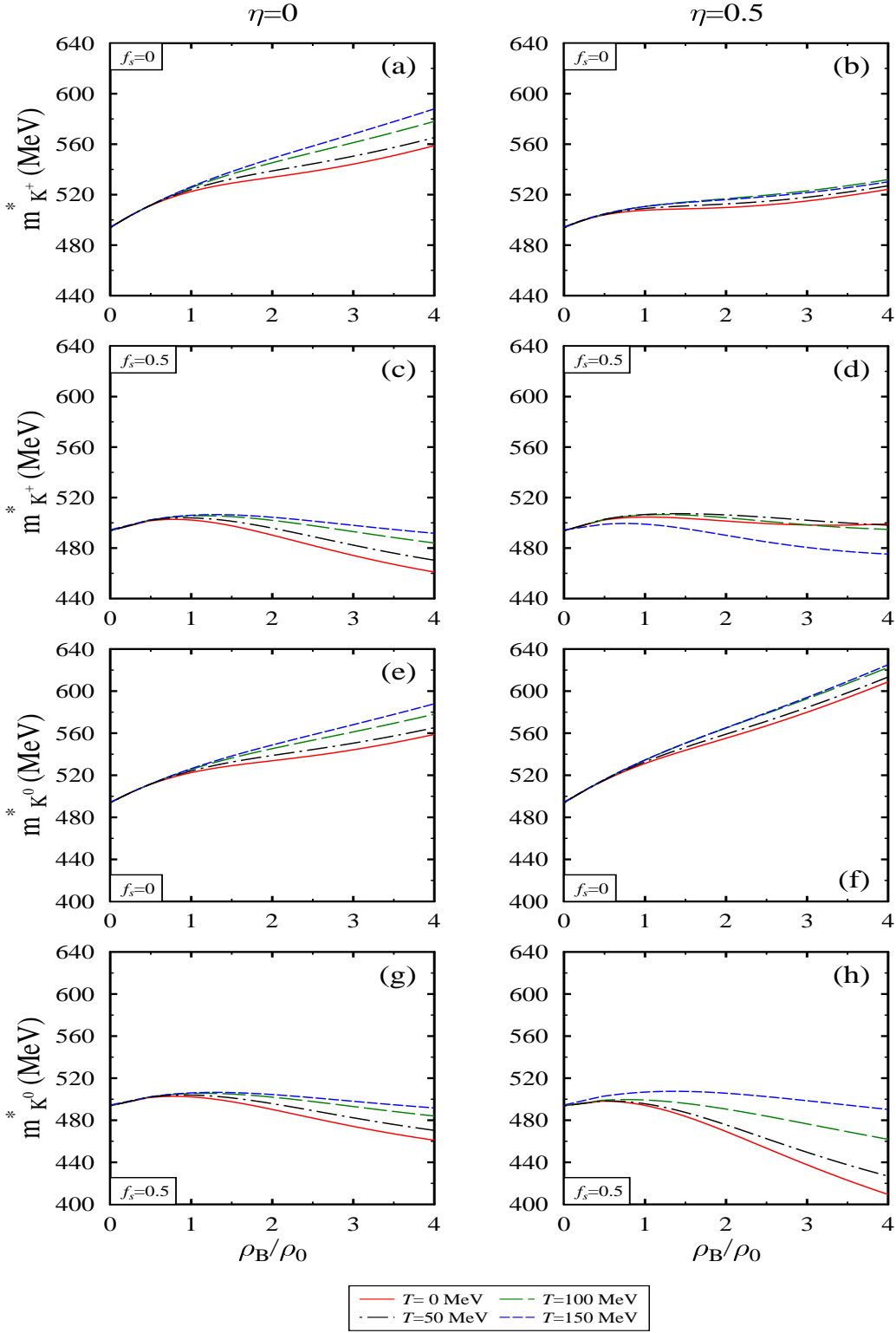}
\caption{(Color online) The in-medium mass of isospin doublet $(K^+,K^0)$ in nuclear and hyperonic matter. }
\label{ms_k}
\end{figure}

In this subsection, we discuss the numerical observations of the kaon and antikaon mass in the strange hadronic matter. We use the medium induced scalar and vector densities of baryons in the dispersion relation (Eq.(\ref{drk})) to calculate the in-medium mass of these mesons. In \cref{tablems}, we listed the in-medium masses of the kaons and antikaons for distinct medium parameters.
 We plot the in-medium mass of $K^+$ and $\bar K^0$ as a function of baryonic density in \cref{ms_k}. We plot this figure for different values of asymmetry, strangeness, and temperature. In a non-strange medium, the mass of  $K^+$ and $K^0$ mesons increase with the increase in baryonic density. It increases almost linearly for high temperature but comparative slow for lower temperatures. As discussed earlier, the various terms of the Eq.(\ref{lagd}) explain the kaon and antikaon baryon interaction and the first term of the Eq.(\ref{lagd}) (known as Weinberg-Tomozawa term) gives repulsive  contributions to the masses of $K^+$ and $K^0$  mesons \cite{Mishra2009,Mishra2008}. The  meson exchange term arising from the  $\sigma$ and $\delta$ fields is attractive for both $K^+$ and $K^0$ mesons.  In isospin symmetric matter ( at a particular value of $f_s$), the  $K^0$ mass shows the exact same behaviour as of $m^*_{K^+}$. This is because the $K^+$ and $K^0$ meson belong to the same isospin doublet. However, in asymmetric nuclear matter, the masses of these mesons do not remain same  because of the asymmetric terms ($\rho_i-\rho_j$; $i\neq j$) present in the Weinberg-Tomozawa term and  isospin dependent range ($d_2$) terms. The self-energy of $K^+$ meson largely depends upon the scalar and vector densities of the baryons which have  positive  $\tau_3$ values, whereas the self-energy of $K^0$ largely depends upon the densities of baryons with negative $\tau_3$ values and due to this  the Weinberg-Tomozawa term  become more repulsive  for $K^0$ mesons and suppress the attractive contributions from $d_2$ term.  When we move from non-strange to strange medium, the mass of $K^+$ meson first slightly increases and then decreases concerning baryonic density for all temperatures.    This is due to the fact that when we go from non-strange to strange medium, the range term, $d_1$ becomes more negative whereas the
$d_2$ term becomes less negative. Therefore the attractive nature of
$d_1$ term dominates the range terms and overall due to this  the mass of $K$ and $K^0$ decreases in the strange medium.  The temperature effects on  the mass of $K^+$ and $K^0$ mesons become less for non-strange asymmetric nuclear matter and for the strange asymmetric matter, it shows appreciable modifications. In the regime of high density, the mass of $K$ mesons decrease with the decrease in temperature. The above behaviour  can be explained on the basis of the temperature dependence of baryon scalar densities. As we increase the medium's temperature, the attractive contributions from range terms start decreasing.

\begin{figure}
\includegraphics[width=16cm,height=21cm]{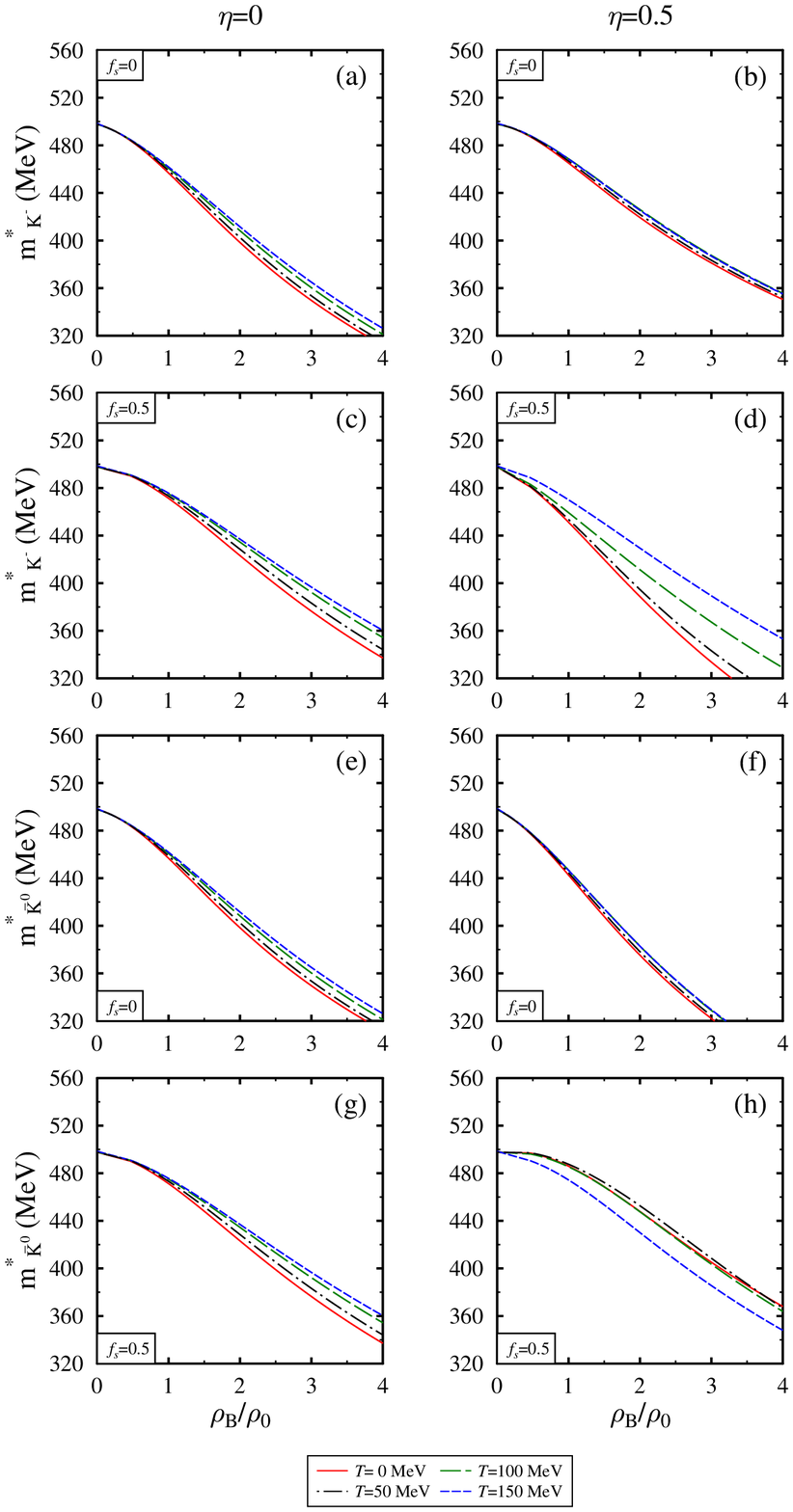}
\caption{(Color online)  The in-medium mass of isospin doublet $(K^-,\bar K^0)$ in nuclear and hyperonic matter. }
\label{ms_kb}
\end{figure}

 In a similar manner, we plot the in-medium mass of antikaons $K^-$ and $\bar K^0$ in \cref{ms_kb}. For the non-strange medium, we observe an opposite behavior in the mass of antikaons as compared to kaons because of the attractive contributions from the Weinberg-Tomozawa term. Furthermore, since $K^-$ and $\bar K^0$ meson also belong to  isospin doublet, the mass of these mesons do not show any difference in the symmetric nuclear matter (for a particular value of strangeness).  The mass of $K^-$ and $\bar K^0$ mesons show less impact of temperature in non-strange but asymmetric baryonic matter  but for  strange asymmetric medium, it shows an appreciable variation with temperature. The explanation lies in the same fact as was discussed for $K$ mesons in the previous paragraph.
 Using the quark meson coupling model,  in Ref. \cite{Martinez2016}, the mass of $K$ meson is studied in non-strange symmetric nuclear matter at zero temperature. In this article, the authors observed that the mass of kaons decreases as a function of baryonic density.  Moreover, in Ref. \cite{Li1995}, using relativistic transport model Li $et.al.$   studied the mass of $K(\bar K)$ meson along with the mass of $\rho$ and $\phi$ meson in the hadronic medium which contains nucleons, pions, and deltas. They observed that the mass of kaons increases with the increase in baryonic density while the  antikaons mass decreases. Using kaon and antikaon in-medium mass, one can also calculate the in-medium optical potential for finite momentum situations via relation ($U^*_{K (\bar K)}(\omega, k) = \omega (k) - m_{K (\bar K)}$) \cite{Mishra2008,Mishra2009}. In Ref. \cite{Lutz1998,Ramos2000,Tolos2001,Tolos2006,Lutz2008},  the vacuum antikaon-nucleon scattering amplitudes are obtained  from the coupled-channel techniques by including the  method of partial waves  to calculate the self-energies of the kaons in the hadronic matter   and observed an attractive potential of the range 40-60 MeV.

 \begin{table}
 \begin{tabular}{|c|c|c|c|c|c|c|c|c|c|}
\hline
& & \multicolumn{4}{c|}{T=50 MeV}    & \multicolumn{4}{c|}{T=150 MeV}   \\
\cline{3-10}
&$f_s$ & \multicolumn{2}{c|}{$\eta$=0} & \multicolumn{2}{c|}{$\eta$=0.5 }& \multicolumn{2}{c|}{$\eta$=0}& \multicolumn{2}{c|}{$\eta$=0.5 }\\
\cline{3-10}
&  &$\rho_0$&$4\rho_0$ &$\rho_0$  &$4\rho_0$ & $\rho_0$ &$4\rho_0$&$\rho_0$&$4\rho_0$ \\ \hline 
$ m^{*}_{{K}^+}$& 0&524&565.1&509.5&527&526.2&587.9&510.6&530.17\\ \cline{2-10}
&0.5&503.9&470.3&506.6&498.2&506&491.7&498.9&475.2 \\ \cline{1-10}
$m^*_{{K}^0}$&0&524&565.1&532.7&613.4&526.2&587.9&534.5&625.1\\  \cline{2-10}
&0.5&503.9&470.4&495.64&426.9&506&491.7&506.7&490.3 \\  
 \cline{1-10}
$ m^{*}_{K^-}$&0&458.7&314.9&466.7&352.2&461.8&326.2&468.5&355.1 \\  \cline{2-10}
&0.5&473.2&344.1&453.6&302.1&475.6&360.3&470&353 \\  \hline
$ m^*_{\bar K^0}$&0&458.7&314.9&444.5&283.3&461.8&326.2&446.6&287.7\\  \cline{2-10}
&0.5&473.2&344.1&487.5&366.8&475.6&360.3&474.3&347.96 \\  \cline{1-10}
\end{tabular}
\caption{In above table, we tabulated the values of in-medium masses   of $K^+,K^0,K^-$ and $\bar K^0$ mesons (in units of MeV).}
\label{tablems}
\end{table}

 \subsection{ In-Medium  Mass and Decay Width of $\phi$ Meson}
 \label{subsec:3.3}

\begin{figure}
\includegraphics[width=16cm,height=16cm]{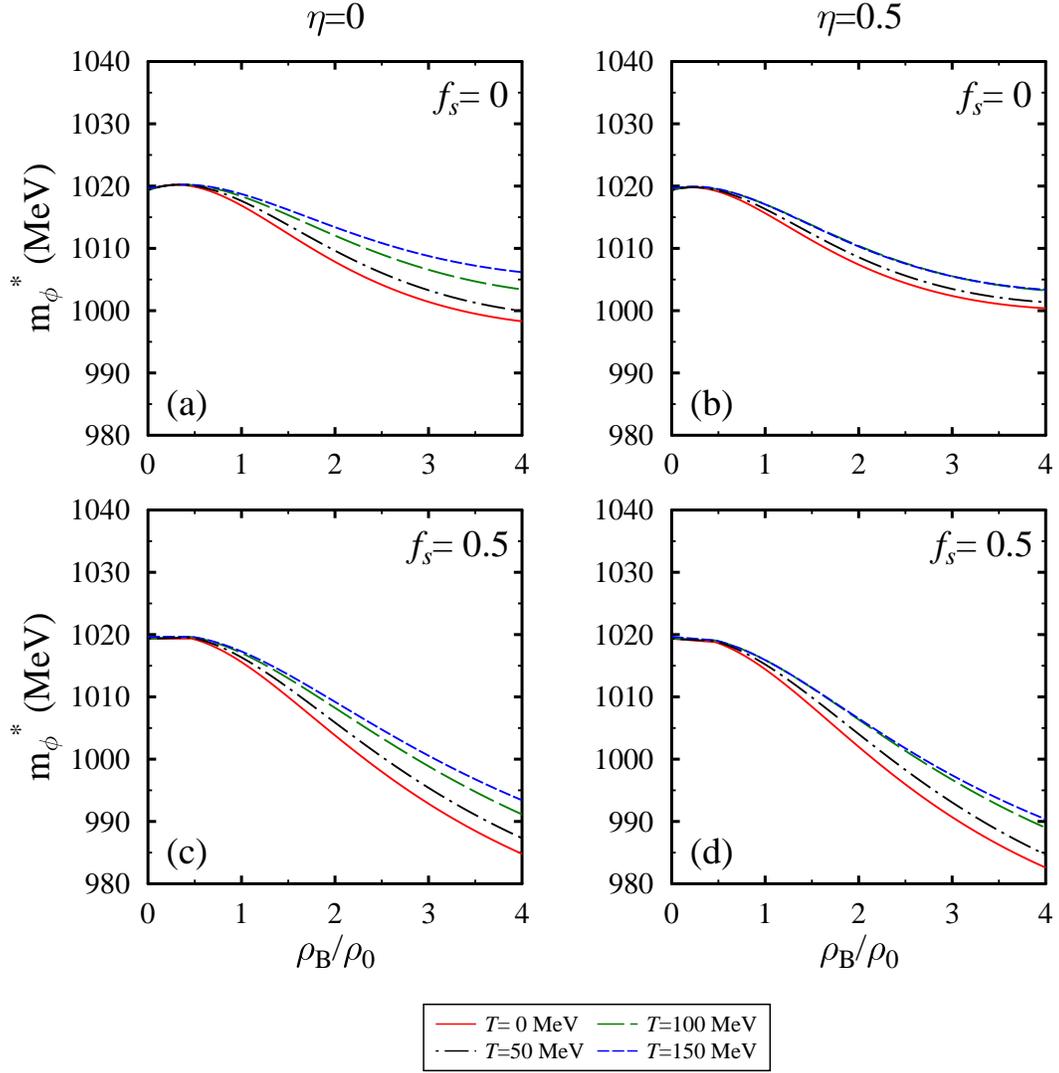}
\caption{(Color online)  The in-medium mass of $\phi$ meson in nuclear and hyperonic matter. }
\label{mphi}
\end{figure}

 The $\phi$ meson mass is calculated through the in-medium self-energy of $\phi$ meson at one loop level (see  subsection \ref{subsec2.3}).  In the previous works of Ref. \cite{Martinez2016,Martinez2017}, the loop integral was solved under the assumption, $i.e.$, $m^{*}_{K }$=$m^{*}_{\bar K}$ but in the present work the self-energy of $\phi$ meson is calculated by solving the regularized loop integral in the presence of medium modified kaons and antikaons where they behave differently as discussed in subsection \ref{subsec:3.2}. Note that the temperature dependence of $\phi$ meson masses and decay width in our present calculations is evaluated through the temperature dependence of kaons and antikaon masses. In \cref{mphi}, we show the in-medium mass of $\phi$ meson as a function of baryonic density by considering the effect of isospin asymmetry, strangeness, and temperature. The medium induced mass for different parameter combinations is  tabulated in the \cref{tabledw1}. From \cref{mphi}, we observe that the mass of $\phi$ meson decreases as a function of medium density. For all permutations of $\eta$ and $f_s$, we see that if the temperature of the medium is decreased then the mass of $\phi$ meson also decreases which reflects the temperature dependence of kaons and antikaons mass. Furthermore, the effect of temperature is more visible in the symmetric matter than the asymmetric matter. Moreover, the increase in strange quarks in the hadronic medium lead to a more decrease in $m^*_\phi$. The in-medium  behaviour of $\phi$ meson mass reflects the in-medium masses of $K$ and $\bar K$ as the self-energy of the $\phi$ meson loop is calculated using these medium modified entities. In Ref. \cite{Martinez2016}, using in-medium $\phi$ self-energy in non-strange symmetric nuclear matter, the medium induced mass of $\phi$ was studied. In this article, Martinez $et.al.$ observed that the mass of $\phi$ meson decrease with the increase of nucleonic density and they plotted the results for three different choices of a cut-off parameter, $\Lambda_c$ $i.e.$ 1,2 and 3 GeV, and found that the $m^*_\phi$ decrease more with the increase in $\Lambda_c$. At $\Lambda_c$=3 GeV and $\rho_B=\rho_0$, they observed 25 MeV decrement in the $\phi$ meson mass whereas we observed 2.59 MeV drop.
   This is because in our calculation kaons and antikaons behave differently in the medium, for example,
   as we discussed earlier, the Weinberg Tomozawa term give the repulsive contribution to $K$ and attractive to $\bar{K}$ meson  mass. 
The values of $m^*_K$ and  $m^*_{\bar K}$ are observed as 522.49 and 456.75 MeV in symmetric nuclear matter at density $\rho_0$ and temperature $T = 0$.   
    However, in Ref. \cite{Martinez2016}, difference in in-medium masses of kaons and antikaons is not taken into calculations and at $\rho_0$, the mass $m^*_K$  is simply considered as $430$ MeV.
 In our article, we observed  less downward shift in $K$ and $ \bar K$ masses and 
therefore a little drop in $\phi$ mass. Using QCD Sum Rules at zero temperature, Klingl $et.al.$, calculated 1 $\%$ drop in $\phi$ mass at nuclear saturation density \cite{Klingl1998}. Furthermore, using the unification of chiral SU(3) model and
 QCD Sum Rules, the author  studied the in-medium mass of $\phi$ meson in asymmetric strange matter at 
zero temperature \cite{Mishra2015} and observed a very small drop. 
They reported the mass shift of about 20 MeV at a density of 5$\rho_0$  in the nuclear medium. 

\begin{figure}
\includegraphics[width=16cm,height=16cm]{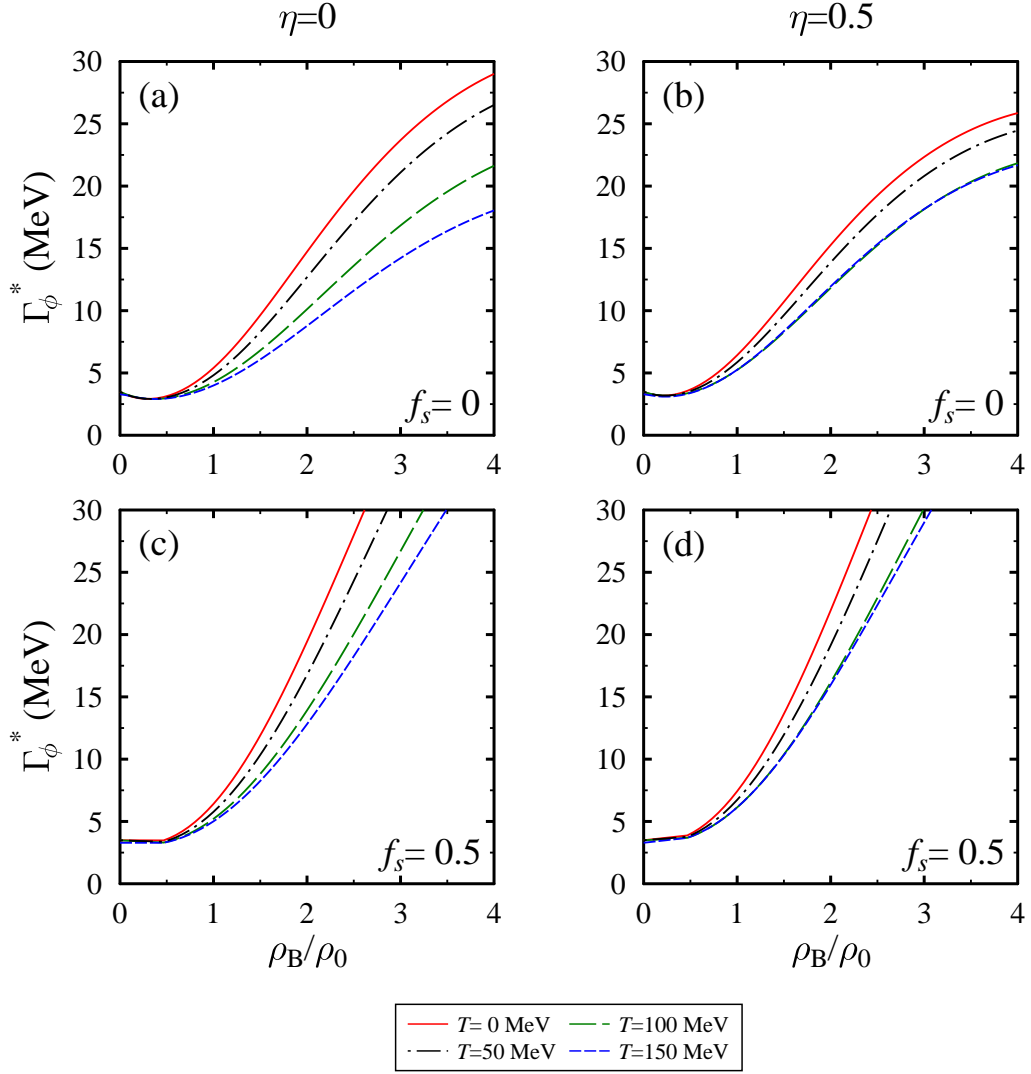}
\caption{(Color online) The in-medium decay width of $\phi$ $\rightarrow$ $K \bar K$ channel
nuclear and hyperonic matter. }
\label{gphi}
\end{figure}

 In \cref{gphi}, we plot the in-medium partial decay width of $\phi$ meson decaying into $K \bar K$ pairs. The formula for decay width was derived by extracting the imaginary part of the  loop integral. The medium modified value of $\Gamma^*_\phi$ is also listed in the \cref{tabledw1} along with $m^*_\phi$.
 In this figure, we observe the partial decay width increase (broadens) with the increase in baryonic density for all cases of strangeness, temperature, and isospin asymmetry. The decay width reflects opposite behavior  as was observed for $m^*_\phi$. However, in the asymmetric matter,  we observe the temperature effects to less appreciable which was also same for $m^*_\phi$. But here in the decay width case, at a particular value of baryonic density the in-medium decay width  increase more for low temperature whereas it was opposite for the mass of  $\phi$ meson. The impact of strange matter with the $\phi$ mesons lead to more increase  in the decay width with baryonic density. This is because of peculiar behavior of $K$ and $\bar{K}$ mesons in strange matter. This highlight the importance of the study of strangeness affects on properties of $\phi$ mesons. We have also calculated  the decay width for different value of $\Lambda_c$ and observed that in strange matter the trend of decay width remains  same concerning baryonic density but its broadening decrease as we lower the value of the cut-off parameter which is consistent with the observations of Ref. \cite{Martinez2016}. As per other theoretical and experimental investigations discussed in the introduction section, the decrement (increment) in $\phi$ meson mass (decay width) in the present investigation is consistent with the observations of existing literature with some distinctions. The cause for these distinctions may lie in the  estimation of the kaon–antikaon loop contributions from different approaches. On the application side, by utilizing the decay width,
the production of $\phi$ mesons in the $pN$ collisions can be measured \cite{Polyanskiy2011}. The comparison of the experimental data with model calculations will speculate the absorption, production, and momentum dependence of $\phi$ meson in the hadronic medium \cite{Paryev2018}.

\begin{table}
\begin{tabular}{|c|c|c|c|c|c|c|c|c|c|c|}
\hline
& & \multicolumn{4}{c|}{T=50 MeV}    & \multicolumn{4}{c|}{T=150 MeV}   \\
\cline{3-10}
&$f_s$ & \multicolumn{2}{c|}{$\eta$=0} & \multicolumn{2}{c|}{$\eta$=0.5 }& \multicolumn{2}{c|}{$\eta$=0}& \multicolumn{2}{c|}{$\eta$=0.5 }\\
\cline{3-10}
&  &$\rho_0$&$4\rho_0$ &$\rho_0$  &$4\rho_0$ & $\rho_0$ &$4\rho_0$&$\rho_0$&$4\rho_0$ \\ \hline

$m^*_\phi$&0&1017.6&999.9&1016.3&1001.3&1018.7&1006.1&1017&1003.3
\\  \cline{2-10} 
 &0.5&1016.3&987&1015&984&1017&993&1016&990\\  \cline{1-10}

$\Gamma^*_\phi$   &0&4.8&26.5&5.8&24.4&3.9&18&5.2&21.6 \\ \cline{2-10} 
&0.5&5&47.5&6&53.1&5&35.9&6&41.5 \\  \cline{1-10}

\hline

\end{tabular}

\caption{In the above table, we tabulated the values of medium induced $\phi$ meson mass and  partial  decay width (MeV) for  $\phi$ $\rightarrow$ $K\bar K$ process for different parameters of the medium.}
\label{tabledw1}
\end{table}

\section{SUMMARY}
\label{sec:4}

To summarize, using an effective Lagrangian approach we calculated the medium modified mass and decay width of $\phi$ meson by employing in-medium $K$ and $\bar K$ masses from the chiral SU(3) model. We have calculated these properties up to four times the nuclear saturation density and in the medium, we considered nucleons and hyperons as degrees of freedom. At finite temperature, we observed appreciable effect of strangeness on the in-medium $K$ and $\bar K$ mesons. The kaon baryon  interactions in a strange medium lead to a decrease in the $K$ and $ \bar K$ mass. The mass of antikaons decreases more appreciably than kaons in the medium. Despite a significant drop in $K$ and $ \bar K$ mass, we observed a small downward mass-shift in the in-medium mass of $\phi$ meson. The impact of temperature becomes less in asymmetric baryonic medium whereas becoming high in the symmetric medium. On the other hand, in the same medium, the decay width shows broadening and it decreases with the increase of strange content in the medium. In the extension of our work, we will study the  appreciable effects on $\phi$ meson such as strangeness enhancement in the $\phi$-mesic nuclei \cite{Jparc}, absorption and production of $\phi$ meson in hadronic medium \cite{Polyanskiy2011,Paryev2018}. It requires future experimental efforts to understand  the medium induced changes in $\phi$ properties in a strange medium. A more systematic study is intended to study the mass-shift of vector mesons with higher statistics in the J-PARC E16 collaboration \cite{E16}. There is also a proposal at J-Lab (following the 12 GeV upgrade), to study the binding of Helium nuclei with $\phi$ and $\eta$ meson \cite{Jlab}. Furthermore, the $K^+/K^0$ and $K^-/\bar K^ 0$ ratios  for different isospin of the beam and
target is promising observables to study the asymmetry  effects in the  CBM experiment at the future
project FAIR at GSI, Germany and Rare Isotope Accelerator (RIA)
laboratory in USA \cite{Mishra2009}.

\centering
\section*{Acknowledgement}

One of the author, (R.K)  sincerely acknowledge the support towards this work from Ministry of Science and Human Resources Development (MHRD), Government of India via Institute fellowship under National Institute of Technology Jalandhar.

\end{document}